\documentstyle[aps,twocolumn,epsfig]{revtex}

\begin{document}
\draft
\title{Spatial interference of coherent atomic waves by
  manipulation of the internal quantum state
}

\author{C.~Fort, P.~Maddaloni$^*$, F.~Minardi, M.~Modugno, and
  M.~Inguscio} \address{ INFM -- L.E.N.S. -- Dipartimento di Fisica,
  Universit\`a di
  Firenze\\
  L.go E.~Fermi~2, I-50125 Firenze, Italy\\
$^*$ Dipartimento di Fisica, Universit\'a di Padova\\
via F.~Marzolo~8, I-35131 Padova, Italy}

\maketitle

\begin{abstract}
  A trapped $^{87}$Rb Bose-Einstein condensate is initially put into a
  superposition of two internal states.  Under the effect of gravity
  and by means of a second transition, we prepare two vertically
  displaced condensates in the same internal state.  These constitute
  two coherent sources of matter waves with adjustable spatial
  separation. Fringe patterns, observed after free expansion, are
  associated with the interplay between internal and external degrees
  of freedom and substantially agree with those for a double slit
  experiment.

\end{abstract} 

Coherent matter waves are available since the achievement of
Bose-Einstein condensation (BEC) in dilute atomic gases
\cite{varenna}. The analogy with coherent light waves was soon
recognized and new perspectives were opened in the fields of atom and
quantum optics \cite{ericelibro}. Interferograms have been reported
\cite{ketterle_int} from two horizontally elongated condensates
separated by a vertical light barrier.  Spatial coherence in the
framework of atom lasers has been investigated by means of radio-frequency
outcoupling \cite{ketterleAL} from selected positions in the
condensate \cite{esslinger}. Bragg diffraction has been used for the
implementation of Mach-Zehnder interferometers with BEC's
\cite{kuga,Simsarian00}. Four-matter waves mixing
\cite{phillipsnature} has opened the field of non-linear atom optics. All
the above schemes are based on the manipulation of the external
degrees of freedom, i.e. position or momentum.  

However, an atomic matter wave also offers the opportunity
to manipulate the internal degrees of freedom. For instance,
coherent superpositions of multistate condensates can be created by
coupling to an external electro-magnetic field.  Using two-photon
pulses coupling, the group at JILA \cite{HallPRL81} has been able to
prepare and investigate two condensates in different hyperfine levels
with a well defined relative phase. This has led to the production of
vortices \cite{MatthewsPRL83} and has allowed the study of a rich
phenomenology. The key physical aspect underlying these experiments
is indeed the intriguing interplay between internal and external
degrees of freedom.  In particular, the coherent preparation of
internal states has been used to alter the topological properties of a
Bose-Einstein condensate
\cite{WilliamsPRA,Nature401,MatthewsPRL83_3358,Hallerice}.
 
In this Letter, we report on a novel coherent matter wave
interferometer that is based on the entanglement between internal and
external degrees of freedom of a condensate. We start with a
magnetically trapped condensate and we create a coherent superposition
of two Zeeman states having different magnetic moments.  The external
degrees of freedom of the two states are differently affected by the
external potential (magnetic field~+~gravity) and the system evolves
towards two spatially separated condensates.  These can be regarded as
two sources of matter waves. After putting them back into the same
quantum state and after free expansion, in the overlapping region the
density distribution is seen to exhibit a clear fringe pattern. The
interferogram is related to the interplay between internal and
external degrees of freedom.  The fringe spacing can be determined by
controlling the evolution time of the quantum state superposition
under the effect of gravity, allowing a quantitative analysis of the
experimental observation.

We produce a $^{87}$Rb condensate of typically $2 \times 10^5$ atoms
in the $F=2, m_F=2$ state trapped in a magneto-static field. The
magnetic trap has a cylindrical symmetry, the axial direction lying in
the horizontal plane \cite{EPJnostro}. The magnetic potential is
harmonic around its minimum. For atoms in the $F=2, m_F=2$ state the
axial frequency is $\omega_{x2}=2 \pi \times 12.6(2)$~Hz and the
radial one is $\omega_{\perp 2}=2 \pi \times 192$~Hz.  As recently
shown \cite{ericenoi,PRL} we can control the transfer of some
population to the other Zeeman sublevels of the $F=2$ state by
applying a resonant radio-frequency (rf) pulse. In the present
experiment we choose the intensity and the length of the rf-pulse so
as to equally populate the two low-field seeking sublevels $m_F=2$ and
$m_F=1$.  Under this condition the untrapped $m_F=0$ sublevel is
populated with 15\% of the total population, while the high-field
seeking sublevels ($m_F=-1,-2$) are essentially unpopulated. As a
consequence we can restrict to a two-level system, neglecting the
$m_F=0,-1,-2$ state and considering the condensate after the rf-pulse
in a superposition of two atomic states. The $m_F=1$ component has
half the magnetic moment of the $m_F=2$ experiencing an harmonic potential
characterized by lower frequencies ($\omega_1=\omega_2 / \sqrt{2}$).
Also, the two potentials, due to the combined effect of magnetic field
and gravity, have different minima vertically separated by
$g/\omega_{ \perp 2}^2$, where $g$ is the acceleration of gravity.  The $m_F=1$
component, which immediately after its production, is in the
equilibrium position for the $m_F=2$ component, starts to oscillate
\cite{PRL}.

We then apply a second rf-pulse after a delay time $t_1$ which can be
varied in the range $0.5 \div 2.2$~ms. The effect of this second pulse
is to produce two couples of vertically displaced wavepackets each in
the same Zeeman state.  We leave the system evolving in the trap
for an additional time $t_2$ before switching off the magnetic field.
During the expansion, a micro-wave pulse is used to transfer to $F=1$
level either $m_F=1$ or $m_F=2$ condensates. After $28$~ms we take an
absorption picture detecting the column density distribution of the
condensates remained in $F=2$.

\begin{figure}

\epsfig{file=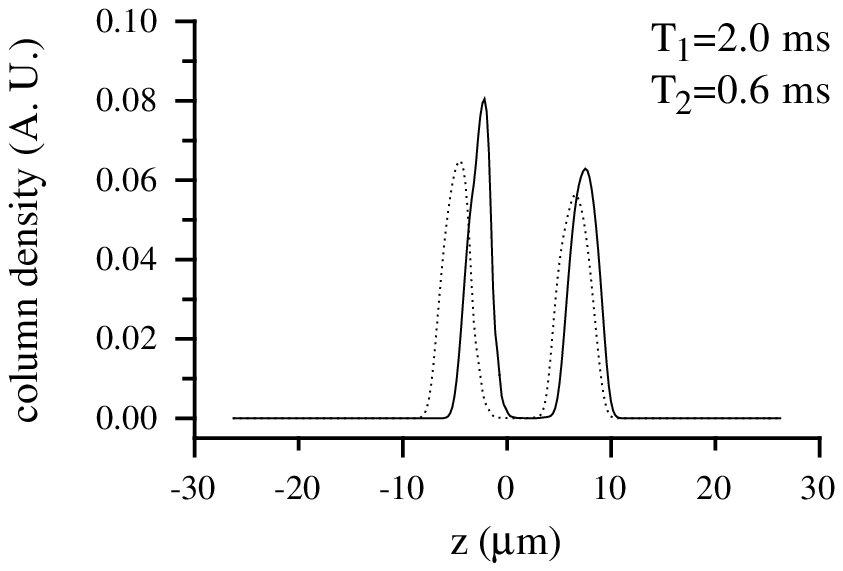,width=5cm}
\epsfig{file=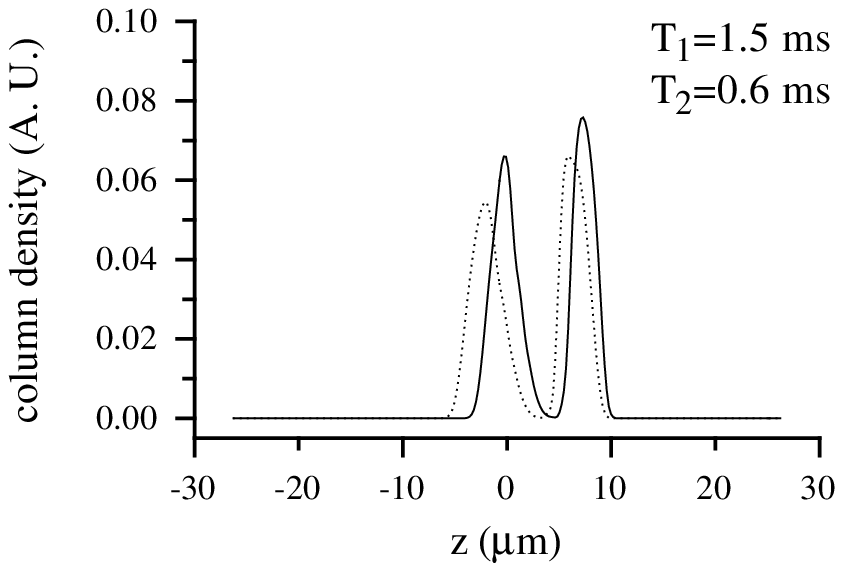,width=5cm}
\epsfig{file=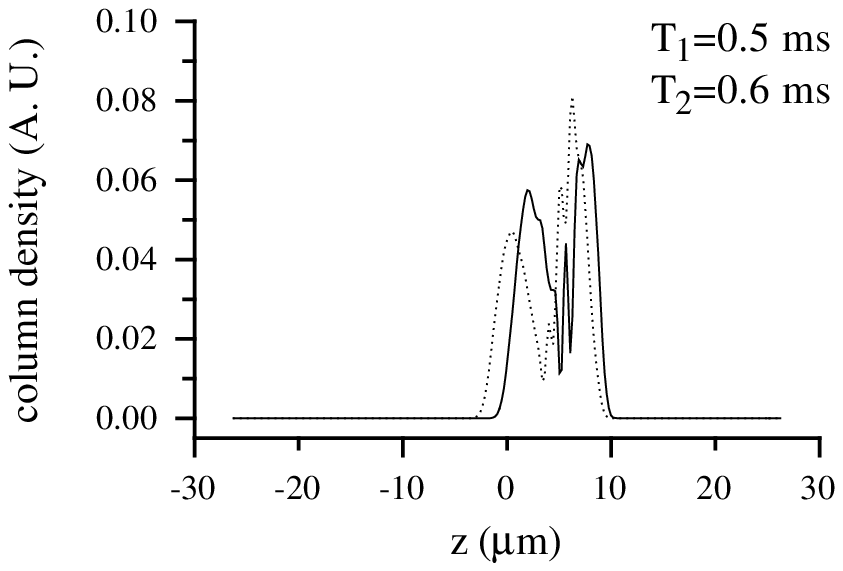,width=5cm}

\caption{Calculated density distributions in $m_F=2$ (continuous line)
  and $m_F=1$ (dotted line) state for three different evolution times
  $t_1+t_2$ in the trap. Properly choosing the timing we can produce
  two completely spatially separated sources of coherent matter
  waves.}
\label{density}
\end{figure}

\begin{figure}
\epsfig{file=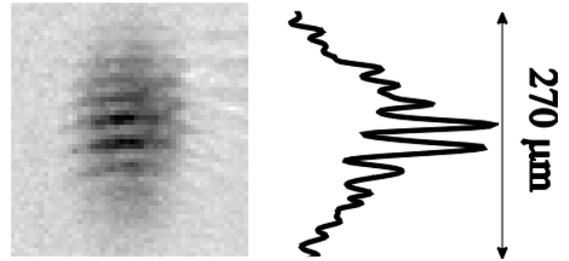,width=8cm}
\caption{Column density distribution resulting from the interference
  of two $m_F=2$ condensates overlapping after 28~ms of free
  expansion. The two condensates are produced in the magnetic trap:
  the time distance between the two rf-pulses is $t_1=1.5$~ms and the
  permanence time in the trap after the second pulse is
  $t_2=0.6$~ms. The curve on the right is a vertical cross-section of
  the recorded image clearly showing spatial fringes.}
\label{frange}
\end{figure}

The dynamics of the two component condensates in the trap, can be
obtained from a two coupled Gross-Pitaevskii equations model
\cite{PRAnostro}.  The vertical separation of the two density
distributions when we switch off the trap depends on gravity through
the time $t_1+t_2$, as shown in Fig.~\ref{density}.  Spatially
separated sources, reminding of a ``double-slit'' configuration, can
be produced either in $m_F=2$ or in $m_F=1$.  By properly choosing the
$t_1$, $t_2$ and the expansion time $T$ we are able to take picture
when the two matter waves overlap and to detect interference fringes
which are shown in Fig.~\ref{frange}.  Unlike photons, interactions
may cause different regimes ranging from partially overlapped to
completely separated atomic clouds. In the latter case one can
consider the system composed by two distinct condensates each in a
state $|\Psi \rangle = \psi(Z) \exp (i \Phi(Z))$ with phases $\Phi(Z)$
which evolve independently from one another during the expansion.  The
vertical evolution of the phase of each expanding condensate is given
by \cite{Simsarian00,dalfovo}
\begin{equation}
\Phi(Z)= \frac{m}{2 \hbar} \frac{\omega_{\perp}^2
  T}{1+\omega_{\perp}^2 T^2}
  Z^2+ \frac{m
  V}
{\hbar}Z
\label{fase}
\end{equation}
where $m$ is the rubidium mass, $\omega_{\perp}$ is the trap frequency
in the vertical direction, $T$ is the expansion time and $V$ is the
condensate center of mass velocity.  The quadratic term in
$Z$  describes the mean-field expansion in the Thomas-Fermi
limit, while the linear term in $Z$ takes into account the center of mass
motion.  For long enough expansion times (in our experiment $\omega^2
T^2 \sim 1000$) Eq.~\ref{fase} reduces to
\begin{equation}
\Phi(Z)= \frac{m}{2 \hbar T} Z^2+ \frac{mV}{\hbar}Z
\end{equation}

\begin{figure}
\epsfig{file=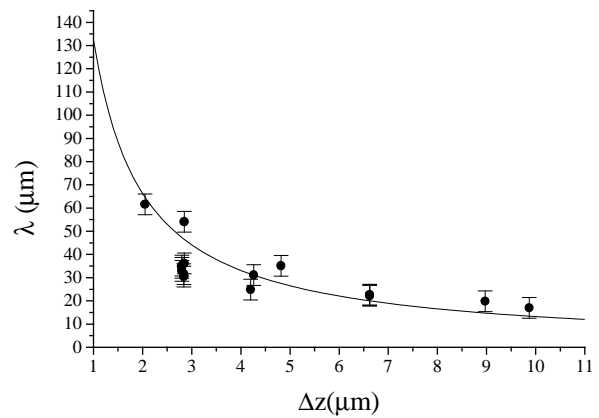,width=8cm}
\caption{Measured vertical fringes spacing in the column density taken
  after $28$~ms of expansion as a function of the vertical
  displacement before expansion between the two interfering
  wavepackets. The solid curve represent the fringes wavelength in a
  double slit experiment (Eq.\protect\ref{teoria}).}
\label{lambda}
\end{figure}

When detection takes place, if the two clouds are slightly displaced
by the amount $\delta Z=Z_1 -Z_2$, one gets an interference term
proportional to
\begin{eqnarray}
\cos(\Phi (Z)-\Phi(Z-\delta Z))=&& \nonumber\\
\cos[(\frac{m}{\hbar T}
(Z_2-Z_1)+ 
\frac{m}{\hbar}(V_1-V_2))Z+C] &&
\end{eqnarray}
where $C$ is a constant phase and $Z_i$, $V_i$ ($i$=1,2) are the
center of mass position and velocity of $i$-th condensate at the
detection time. The fringes spacing is simply
\begin{equation}
\lambda=\frac{h T}{m}\left( \frac{1}{(Z_2-Z_1)+T(V_1-V_2)}\right)
\label{lZV}
\end{equation}
and we could in principle measure the relative velocity knowing
$(Z_2-Z_1)$. Furthermore, from the free-fall equations, it is
straightforward to obtain
\begin{equation}
\lambda= \frac{hT}{m(z_2-z_1)}
\label{teoria}
\end{equation}  
where $\lambda$ depends only on the distance $z_2-z_1$ between the two
condensates before the expansion like for a double slit.  The
predictions from Eq.~\ref{teoria} together with data corresponding to
different ``double-slit'' separations are reported in
Fig.~\ref{lambda}. The agreement is very good for large $\Delta z$'s,
while it continues to be satisfactory for $\Delta z <5$~$\mu$m
corresponding to condensates not completely separated in which case
interactions could play an important role \cite{wallis,walls}.

In conclusion, we have implemented a novel matter wave interferometer
based on the control of the interplay between internal external
degrees of freedom.  The separation of a double slit BEC can be
adjusted by simply varying the pulse duration and the delay time.
Interferograms yield information on the acceleration of gravity as
well as on the relative velocity of the two interfering matter
wavepackets.  In addition, future developments of this matter waves
interferometer would allow to study the regime of overlapping
condensates where, unlike photons, new effects are expected to arise
from mutual interactions between the sources.

This work benefitted of stimulating discussion with S.~Stringari and
F.~Dalfovo and was supported by the Cofinanziamento MURST and by the
CEE under Contract No. HPRICT1999-00111. The authors would like to
thank M.~Artoni for critical reading of the manuscript.



\end{document}